# *Basis of Local Approach in Classical Statistical Mechanics.*

## S. R. Sharov

Tomsk, Russia e-mail: sharov-s@yandex.ru

An ensemble of classical subsystems interacting with surrounding particles has been considered. In general case, a phase volume of the subsystems ensemble was shown to be a function of time. The evolutional equations of the ensemble are obtained as well as the simplest solution of these equations representing the quasi-local distribution with the temperature pattern being assigned. Unlike the Gibbs's distribution, the energy of interaction with surrounding particles appears in the distribution function, which make possible both evolution in the equilibrium case and fluctuations in the non-equilibrium one. The expression for local entropy is obtained. The derivation of hydrodynamic equations from Boltzmann equation has been analyzed. The hydrodynamic equations obtained from Boltzmann equation were shown to be equations for ideal liquid. The exact expressions for changing entropy and quantity of the heat given by the environment have been obtained. A two-particle distribution function for pair interaction system has been obtained with the use of local conditional distribution functions. Its formula is exact disregarding edge conditions. Reasons for stochastic description in deterministic Hamilton's systems, conditions of applicability of Poincaré's recurrence theorem as well as the problem of irreversibility have been considered.

One of the problems of natural science is the problem of establishing a link between mechanics and thermodynamics. For the equilibrium case this problem was solved by Gibbs. Up to the present, however, there is no satisfactory solution in the general case despite a great number of papers on the issue.
Another aspect of the problem is: how can one get a stochastic description in deterministic Hamiltonian systems? An attempt made to answer those and some other questions on the basis of the approach suggested by the author of this paper is given below.

## Mechanics and Ensembles.

The conception of Gibbs's ensemble /1/ is known to be the basis of statistical mechanics, with equations of mechanics being used in the form of Liouville's equation. This differential equation in partial derivatives of the first order is equivalent to Hamilton's equations. Still the conception of Gibbs′s ensemble, which proved valid in the equilibrium case, does not lead to satisfactory results in an attempt to generalize it for the general case. No one has succeeded in finding a non-trivial solution of Liouville's equation up to now. And we think the reason for that is not the weakness of mathematical methods, but actual nonstationarity of nonequilibrium process in general case, which cannot be ergodic. Moreover, a possibility to give a correct definition of entropy for a nonequilibrium case is doubtful. That Gibbs's definition of entropy fails to suit for general case, is clearly seen from considering the distribution function for a canonical ensemble, since temperature of the system in question /1/ enters the formula of Gibbs's entropy in terms of this distribution function. However, in the nonequilibrium case the system can have different temperatures in its diverse parts or it can even be characterized by a temperature field.

The application of Gibbs's ensemble to a part of the system is invalid by the following reason: in constructing the Gibbs's ensemble, the coordinates of external bodies are considered to be equal (and macroscopical) for all the ensemble forming systems. In this case, the system can only exchange work with its environment but not heat, so it is adiabatic. The Gibbs's ensemble, therefore, is not the most general one, and a view about the canonical Gibbs's distribution as describing the system in the thermostat is not exactly correct. The nature of this discrepancy will be elucidated later. Taking into account the above, it is reasonable from the very beginning to construct the classical statistical mechanics on the conception of ensemble which is more general than the Gibbs's ensemble. Such an ensemble should consist of systems, more exactly of subsystems, exchanging with the environment not only work but also heat.

For this purpose, it is necessary and sufficient to assume that coordinates and momenta of external bodies, i.e. particles surrounding the subsystem under consideration, might have different values for each of the subsystems. Any nonequilibrium thermodynamic system can be presented as consisting of such subsystems being infinitely small in general case. It is implied that all the systems are under similar macroscopical conditions which can be characterized by assigning fields of thermodynamic parameters. If we manage to define the entropy for any subsystem so that it could change owing to exchange of heat with the surrounding subsystems, then the entropy of the whole system can change in integral considering the whole system, similarly to the case with thermodynamic consideration. Hereinafter such an ensemble is referred to as the ensemble of subsystems. Thus, let us consider the ensemble of subsystems and define the equations for its evolution to be described.

To construct the ensemble of subsystems, let us consider any thermodynamic system, whose state is assumed to be given "macroscopically", and take conceptually an infinite number of its copies characterized by the same thermodynamic parameters. Let each system consist of $N$ particles. Mentally single out a subsystem of $n$ particles $(1 \le n < N)$ from every system and represent it as a point in the phase space of dimension $6n$. In this case we consider coordinates and momenta of external (for the subsystem) particles as assigned, though unknown. Generally speaking, they will be different for each subsystem. With this, the phase volume occupied by the singled out set of phase points in the space of dimension $6n$, will change with time. At first glance, this statement is contrary to the well known theorem of the phase volume conservation by Hamiltonian systems.

Actually there is no contradiction, and the use of the mentioned theorem does make the result to be almost obvious. To understand this, one should otherwise approach the construction of the ensemble of subsystems: it may be grasped that the above ensemble presents a projection of section of Gibbs's ensemble of dimension $6N$ by a hypersurface, on a subspace of dimension $6n$. But it in no way follows that the projection area of the singled out part of the section will not change with time

We now proceed to analytical consideration of the problem. Consider the ensemble of subsystems, the evolution of each of them being defined by Hamilton's equations:

$$\dot{q}_\alpha = \frac{\partial H}{\partial p_\alpha}; \quad \dot{p}_\alpha = -\frac{\partial H}{\partial q_\alpha} \quad 1 \le \alpha \le 3n \tag{1}$$

where $q_\alpha$ is $3n$ of generalized coordinates,

$p_\alpha$ is $3n$ of generalized momenta

$H = H(q_1, \cdots q_{3N}, p_1, \cdots p_{3N}, t)$ is the Hamiltonian of the whole system, which depends on $3N$ coordinates and momenta of all $N$ particles and, probably, on time $t$.

Now we are to show that

$$\int \ldots \int dq_1^0 \ldots dq_{3n}^0 dp_1^0 \ldots dp_{3n}^0 \ne \int \ldots \int dq_1 \ldots dq_{3n} dp_1 \ldots dp_{3n}$$

that is the phase volume calculated at the time point $t^0$ is not equal to the corresponding phase volume calculated at the time point $t \ne t^0$. To do this, it would seem possible to use a well-known technique of change from integration with respect to one set of variables to integration with respect to another one using the Jacobian

$$J = \frac{\partial(q_1, \cdots p_{3n})}{\partial(q_1^0, \cdots p_{3n}^0)}$$

and to write with denoting the multiple integration by one symbol $\int$

$$\int dq_1 \ldots dp_{3n} = \int |J| dq_1^0 \ldots dp_{3n}^0 \tag{2}$$

To use this formula, however, it is necessary that one-to-one correspondence would exist between the positions of the phase point at moments $t$ and $t^0$.

In this case, generally speaking, there is no such a correspondence (which requires a change to a statistical description by itself). Nevertheless, such a correspondence is certain to take place at start time and $J_0 = 1$, i.e. formula (2) holds for the case of subsystems for infinitesimal transformations. Let us show that $dJ/dt$ is not equal to zero identically. To make it obvious, first we consider the simplest case where the initial system involves only two particles executing one-dimensional motion. And we take one particle as the subsystem. We are going to show that

$$\int dq_1^0 dp_1^0 \ne \int dq_1 dp_1$$

Differentiate $J$ with respect to time

$$\frac{dJ}{dt} = \begin{vmatrix} \frac{\partial \dot{q}_1}{\partial q_1^0} & \frac{\partial \dot{q}_1}{\partial p_1^0} \\ \frac{\partial p_1}{\partial q_1^0} & \frac{\partial p_1}{\partial p_1^0} \end{vmatrix} + \begin{vmatrix} \frac{\partial q_1}{\partial q_1^0} & \frac{\partial q_1}{\partial p_1^0} \\ \frac{\partial \dot{p}_1}{\partial q_1^0} & \frac{\partial \dot{p}_1}{\partial p_1^0} \end{vmatrix} \tag{3}$$

Using Hamilton's equations:
$$\dot{q}_1 = \frac{\partial H(q_1, p_1, q_2, p_2)}{\partial p_1}; \quad \dot{p}_1 = -\frac{\partial H(q_1, p_1, q_2, p_2)}{\partial q_1}$$
one can write
$$\frac{\partial \dot{q}_1}{\partial q_1^0} = \frac{\partial \dot{q}_1}{\partial q_1}\frac{\partial q_1}{\partial q_1^0} + \frac{\partial \dot{q}_1}{\partial p_1}\frac{\partial p_1}{\partial q_1^0} + \frac{\partial \dot{q}_1}{\partial q_2}\frac{\partial q_2}{\partial q_1^0} + \frac{\partial \dot{q}_1}{\partial p_2}\frac{\partial p_2}{\partial q_1^0} \quad (4)$$

and analogous expressions for $\frac{\partial \dot{q}_1}{\partial p_1^0}; \frac{\partial \dot{p}_1}{\partial q_1^0}; \frac{\partial \dot{p}_1}{\partial p_1^0}$

In expression ( 4 ) the presence of the third and fourth terms is very essential, which contain partial derivatives with respect to $q_2$ and $p_2$, i. e. with respect to coordinates and momenta of surrounding particles. When substituting ( 4 ) into ( 3 ), either of determinants is divided into the sum of four determinants, with determinants free of partial derivatives with respect to $q_2$ and $p_2$ being equal to zero (similarly to the case of Gibbs's ensemble /1/), and combining the retaining terms we obtain:

$$\frac{dJ}{dt} = \begin{vmatrix} \frac{\partial \dot{q}_1}{\partial q_2}\frac{\partial q_2}{\partial q_1^0} + \frac{\partial \dot{q}_1}{\partial p_2}\frac{\partial p_2}{\partial q_1^0}, & \frac{\partial \dot{q}_1}{\partial q_2}\frac{\partial q_2}{\partial p_1^0} + \frac{\partial \dot{q}_1}{\partial p_2}\frac{\partial p_2}{\partial p_1^0} \\ \frac{\partial p_1}{\partial q_1^0} & \frac{\partial p_1}{\partial p_1^0} \end{vmatrix} + \begin{vmatrix} \frac{\partial q_1}{\partial q_1^0} & \frac{\partial q_1}{\partial p_1^0} \\ \frac{\partial \dot{p}_1}{\partial q_2}\frac{\partial q_2}{\partial q_1^0} + \frac{\partial \dot{p}_1}{\partial p_2}\frac{\partial p_2}{\partial q_1^0}, & \frac{\partial \dot{p}_1}{\partial q_2}\frac{\partial q_2}{\partial p_1^0} + \frac{\partial \dot{p}_1}{\partial p_2}\frac{\partial p_2}{\partial p_1^0} \end{vmatrix}$$

One can write the obtained results by convention in the form:
$$\frac{dJ}{dt} = \frac{d}{dt}\frac{\partial(q_1\{q_2, p_2\}, p_1\{q_2, p_2\})}{\partial(q_1^0, p_1^0)} \quad (5)$$

This notation means merely that differentiation of the kind $\frac{\partial \dot{q}_1}{\partial q_1^0}$ is made by way of intermediate variables $q_2$ and $p_2$ while the variables $q_1$ and $p_1$ are lost. It is seen that in general case $\frac{dJ}{dt} \neq 0$

Similarly, one can consider a more general case with systems of $N$ particles and subsystems of n-particles. Calculations become more unwieldy, but the final result has the same form
$$\frac{dJ}{dt} = \frac{d}{dt}\frac{\partial(q_1\{q_\kappa; p_\kappa\}, \cdots p_{3n}\{q_\kappa; p_\kappa\})}{\partial(q_1^0, \cdots p_{3n}^0)} \neq 0 \quad (6)$$
where from here on indices $k$ mark the numbers of coordinates and momenta of the system particles which don't belong to the subsystem under consideration. The meaning of the result consists in the fact that a contribution to changes of phase volume is made by those changes of $q_\alpha$ and $p_\alpha$ of coordinates and momenta of the subsystem particles, which are caused by interactions of the subsystem particles with environmental particles. It is obvious that in the limiting case of absence of subsystems interaction with the environment, we obtain the Gibbs's ensemble and the phase volume conservation. One can note that it is seen from equations ( 5 ) and ( 6 ) that a phase volume change for subsystems is not directly related to velocity divergence (see in more detail hereinafter). Moreover, from the above one can easily obtain that for the subsystems
$$\frac{\partial(q_1, \cdots p_{3n})}{\partial(q_1^0, \cdots p_{3n}^0)} \neq \frac{\partial(q_1, \cdots p_{3n})}{\partial(q_1', \cdots p_{3n}')} \times \frac{\partial(q_1', \cdots p_{3n}')}{\partial(q_1^0, \cdots p_{3n}^0)}$$
where $q_\alpha'$ are the coordinates at the time point $t'$. This proves again that equation ( 2 ) is inapplicable in this case.

Introduce now the distribution function $\rho_n$ for the ensemble of subsystems of $n$ particles similarly to the case of Gibbs's statistics. This function defining probability of values of coordinates $q_\alpha$ and momenta $p_\alpha$ of $n$ subsystem particles can also depend on time $t$ and parametrically on coordinates $q_\kappa$ and momenta $p_\kappa$ of particles surrounding the subsystem. Therefore the total derivative of the distribution function $\rho_n$ with respect to time should be written in the form:

$$\frac{d\rho_n}{dt} = \frac{\partial \rho_n}{\partial t} + \sum_{\alpha=1}^{3n}\left(\frac{\partial \rho_n}{\partial q_\alpha}\dot{q}_\alpha + \frac{\partial \rho_n}{\partial p_\alpha}\dot{p}_\alpha\right) + \sum_{\kappa=1}^{3(N-n)}\left(\frac{\partial \rho_n}{\partial q_\kappa}\dot{q}_\kappa + \frac{\partial \rho_n}{\partial p_\kappa}\dot{p}_\kappa\right) \qquad (7)$$

Consider now the possibility of applying a continuity equation in $6n$-space of the ensemble of subsystems. The absence of velocity spread in the phase space is the condition of application the continuity equation, i. e. at the infinitely near space points the velocity values are bound to differ by the infinitesimal /2/. In this case the spreading may appear in the subspace of momenta $p_\alpha$, for the velocity $\dot{p}_\alpha$ in it represents a force acting on the particle. In principle, a situation is possible when at an infinitely small difference in the positions of two points in the $6n$-space, forces acting from the environment differ by a finite quantity due to a finite difference in the positions of surrounding particles. This case, however, can be easily avoided by means of proper choice of subsystems. For this purpose, there must be no discontinuity in the hypersurface intersecting Gibbs's ensemble so as to form an ensemble of subsystems, i. e. the hypersurface must be smooth. This condition is not essentially burdensome and we can use the continuity equation in the $6n$-space as well, where it takes the form

$$\frac{\partial \rho_n}{\partial t} + \sum_{\alpha=1}^{3n}\left\{\frac{\partial}{\partial q_\alpha}(\rho_n \dot{q}_\alpha) + \frac{\partial}{\partial p_\alpha}(\rho_n \dot{p}_\alpha)\right\} = 0 \qquad (8)$$

Using Hamilton's equations and the equality of second mixed derivatives, one can reduce equation (8) to the following form:

$$\frac{\partial \rho_n}{\partial t} + \sum_{\alpha=1}^{3n}\left(\frac{\partial \rho_n}{\partial q_\alpha}\dot{q}_\alpha + \frac{\partial \rho_n}{\partial p_\alpha}\dot{p}_\alpha\right) = 0 \qquad (9)$$

From the form of equation (9), it is bound seemingly to follow $\frac{d\rho_n}{dt} = 0$, for the second member is equal to zero. Still, since $\rho_n$ depends, generally speaking, on $6N$ variables and time, $6N$ equations will be equations of characteristics for it, from them $6n$ equations being of the form

$$\frac{dq_\alpha}{dt} = \frac{\partial H}{\partial p_\alpha}; \quad \frac{dp_\alpha}{dt} = -\frac{\partial H}{\partial q_\alpha} \quad 1 \le \alpha \le 3n$$

and $6(N-n)$ equations of the form

$$\frac{dq_\kappa}{dt} = 0; \quad \frac{dp_\kappa}{dt} = 0 \quad 1 \le \kappa \le 3(N-n)$$

The index $k$ relates to coordinates and momenta of external particles, which are not constant but change according to Hamilton's equations. In view of this:

$$\frac{d\rho_n}{dt} \ne 0$$

Substituting (9) into (7) we obtain

$$\frac{d\rho_n}{dt} = \sum_{\kappa=1}^{3(N-n)}\left(\frac{\partial \rho_n}{\partial q_\kappa}\dot{q}_\kappa + \frac{\partial \rho_n}{\partial p_\kappa}\dot{p}_\kappa\right) \qquad (10)$$

It is appropriate here to make a small digression and say about the relationship between velocity divergence and volume change. It is well-known Liouville's theorem from the theory of differential equations /5/, according to which, with velocity divergence being equal to zero, the conservation of volumes takes place in mapping. This theorem is proved for the case when dimensionality of a volume whose change is considered in mapping, is equal to the number of variables and to the number of differential equations whose solutions perform mapping of the volume. But in this case we consider mapping of section of dimensionality $6n$, which are less than the total number $6N$ of variables and differential equations.

In this case, the above theorem does not already take place. Instead of considering where exactly the conditions of theorem applicability are violated, we illustrate this by means of a simple example.
Let us consider a system of two equations and a one-dimensional ensemble in the X-direction

$$\dot{x} = y; \quad \dot{y} = -x$$

Here y is not a generalized momentum. $\qquad (11)$

A velocity divergence in the X-direction calculated from the first equation of the system ( 11 ) is equal to 0

$$\frac{\partial \dot{x}}{\partial x} = \frac{\partial y}{\partial x} = 0$$

And the general solution of the system ( 11 ) has the form

$$x = x_0 \cos t + y_0 \sin t$$
$$y = -x_0 \sin t + y_0 \cos t$$

Hence, having assumed for simplicity $y_0 = 0$ we see that the length of any segment X-direction is not constant in time. That is in the one-dimensional ensemble under consideration, the volume is not conserved though the velocity divergence is equal to 0.

After this short digression, we turn again to equation ( 9 ), by its form it does not differ from Liouville's equation. There is an important distinction, however, consisting in the fact that terms of the form $\frac{\partial \rho_n}{\partial p_\alpha} \dot{p}_\alpha$ contain forces acting on particles of the subsystem in question from surrounding particles. Those are "dissipative" forces which on averaging, for example, can give forces proportional to velocity acting on Brownian particles, i.e. the considered approach allows one to build a bridge between Hamiltonian and dissipative systems. The same forces cause energy fluctuations in subsystems. Moreover, they are also responsible for other effects that will be considered below.

We now turn to the search of solutions of the equations. First for simplicity, we shall seek stationary solutions, i.e. those satisfying the condition $\frac{\partial \rho_n}{\partial t} = 0$. Assume that Hamiltonian of the whole system does not explicitly depend on the time $t$ and represent it in the form

$$H = H_1(q_\alpha, p_\alpha) + H_2(q_\kappa, p_\kappa) + H_w(q_\alpha, q_\kappa)$$

where $H_1(q_\alpha, p_\alpha)$ is energy of the subsystem under consideration without regard for the environment; $H_w(q_\alpha, q_\kappa)$ is energy of interaction with the environment of the subsystem; $H_2(q_\kappa, p_\kappa)$ is energy of the environment. For simplicity, we have supposed here $H_w$ as independent of momenta, which is not a principal restriction. Let us denote by $E$ the expression of the form

$$E = E(q_\alpha, p_\alpha, q_\kappa) \equiv H_1(q_\alpha, p_\alpha) + H_w(q_\alpha, q_\kappa)$$

Then according to Hamilton's equations

$$\dot{q}_\alpha = \frac{\partial H}{\partial p_\alpha} = \frac{\partial E}{\partial p_\alpha}; \quad \dot{p}_\alpha = -\frac{\partial H}{\partial q_\alpha} = -\frac{\partial E}{\partial q_\alpha} \quad (12)$$

Hence

$$\frac{dE}{dt} = \sum_{\alpha=1}^{3n}\left(\frac{\partial E}{\partial q_\alpha}\dot{q}_\alpha + \frac{\partial E}{\partial p_\alpha}\dot{p}_\alpha\right) + \sum_{\kappa=1}^{3(N-n)} \frac{\partial E}{\partial q_\kappa}\dot{q}_\kappa = \sum_{\kappa=1}^{3(N-n)} \frac{\partial E}{\partial q_\kappa}\dot{q}_\kappa$$

The last equation is obtained in terms of ( 12 ).

Now it is possible to write the simplest equation for the distribution function $\rho_n$ satisfying equations ( 9 ) and ( 10 ) as well as the condition $\frac{\partial \rho_n}{\partial t} = 0$

$$\rho_n = \exp(-E/\Theta) \quad (13)$$

where $\Theta$ is the parameter, the meaning of which (temperature in energy units) will be revealed below.

Since according to the above assumption $\rho_n$ is the distribution function of subsystems in the phase space, then it is subject to normalization to the unit.

$$\int \rho_n dq_1 \cdots dq_{3n} dp_1 \cdots dp_{3n} = 1 \quad (14)$$

where integration is performed over the whole phase space ( $6n$ dim),

In accordance with the said above, we choose the distribution function in the form

$$\rho_n = \exp[(F-E)/\Theta] \quad (15)$$

where $F$ depends on $\Theta$ and $q_\kappa$, but not on $q_\alpha$ and $p_\alpha$. So equations (9) and (10) are satisfied by the solution (15). This solution is a local, more exactly quasilocal distribution in the assigned field of temperatures, that depends on the environment. The environmental dependence enters through the interaction energy $H_w$ which provides fluctuations in an equilibrium case and evolution in a non-equilibrium case.

The limiting transfer to the canonical Gibbs's distribution is obvious. It is also clear that the canonical Gibbs's distribution is merely the limiting case of distribution in a thermostat at $H_w \to 0$

## Entropy

Consider expression (15) in more detail. This distribution function contains coordinates $q_\kappa$ as parameters, i.e. it is a conditional distribution function. But the values of $q_\kappa$ are unknown for us. Nevertheless, we can construct an ensemble for a neighboring subsystem and define probability of values of all the coordinates $q_\kappa$ or a part thereof. It is coordinates $q_\alpha$ of the first subsystem that will now enter into the distribution function as parameters. We can not generally establish such boundary conditions for a subsystem that they should be independent of the subsystem itself. It is not a drawback of this method but presents a more general case of describing nature. Indeed, when we establish boundary conditions "strictly", we introduce certain idealization neglecting a reverse action of the system under consideration on the surroundings. It is clear that such idealization is not always permissible. Since (15) involves the interaction energy of subsystems, which reflects the fact that the subsystems are not statistically independent, then the distribution function of the whole system can not be obtained by factorization of distribution functions of the subsystems. However, with weak coupling being present, such an approximation is possible. Since in this case subsystem temperatures can differ, the total distribution function will not be symmetrical in respect to a rearrangement of coordinates of particles belonging to two different subsystems. Nevertheless, in rearranging particle coordinates within a subsystem, the distribution function will not change, i.e. a local symmetry takes place.

Now turn again to expression (14) and rewrite it taking into consideration (15)

$$e^{-\frac{F}{\Theta}} = \int e^{-\frac{E}{\Theta}} d\Gamma \qquad (16)$$

where $d\Gamma = dq_1 \cdots dq_{3n} dp_1 \cdots dp_{3n}$

and take a variation from both sides of (16) in changing the coordinates $q_\kappa$ and parameter $\Theta$.
In doing this we assume vanishing the distribution function at boundaries of the integration domain:

$$e^{-\frac{F}{\Theta}} \times \left( -\frac{\delta F}{\Theta} + \frac{F}{\Theta^2}\delta\Theta \right) = \frac{1}{\Theta^2}\delta\Theta \int E e^{-\frac{E}{\Theta}} d\Gamma - \frac{1}{\Theta} \int \sum_{\kappa=1}^{3(N-n)} \frac{\partial E}{\partial q_\kappa} \delta q_\kappa e^{-\frac{E}{\Theta}} d\Gamma$$

The expression $-\frac{\partial E}{\partial q_\kappa} \equiv F_\kappa$ represents the force acting on the environment particles by the subsystem. Now we multiply both sides of the obtained expression by $\exp(F/\Theta)$ and use the formula for the average over the ensemble

$$\langle \Psi \rangle = \int \Psi \rho_n d\Gamma$$

where $\Psi$ is an arbitrary function of the coordinate $q_\alpha$ and momenta $p_\alpha$.
Then

$$-\frac{\delta F}{\Theta} + \frac{F\delta\Theta}{\Theta^2} = \frac{1}{\Theta^2}\langle E \rangle \delta\Theta + \frac{1}{\Theta}\sum_{\kappa=1}^{3(N-n)} \langle F_\kappa \rangle \delta q_\kappa$$

Hence

$$\delta F = \frac{F - \langle E \rangle}{\Theta}\delta\Theta - \sum_{\kappa=1}^{3(N-n)} \langle F_\kappa \rangle \delta q_\kappa \qquad (17)$$

Since it follows from (15) that $\ln \rho_n = [(F-E)/\Theta]$
then

$$\delta F = \langle ln\rho_n \rangle \delta\Theta - \sum_{\kappa=1}^{3(N-n)} \langle F_\kappa \rangle \delta q_\kappa \qquad (18)$$

Comparing (18) with the expression for free energy /6/ known from local thermodynamics
$$\delta F = -S\delta\Theta - P\delta V$$
we conclude that $S = -\langle ln\rho_n \rangle$ can be interpreted as local entropy in appropriate units, $\Theta$ is the subsystem temperature. The second term of expression (18) is related to the expansion work $-P\delta V$ where $P$ is pressure, $\delta V$ is the volume variation. A detailed analysis of the expression $-\sum \langle F_\kappa \rangle \delta q_\kappa$ will allow one to understand under what conditions the approximation of local thermodynamics for the expansion work is valid. Turning back to the definition of local entropy $S = -\langle ln\rho_n \rangle$ we see that it is the same in form as the Gibbs's one but has somewhat another content. Now entropy is a dynamical variable changing in accordance with a change of $q_\kappa$ and it is not a strictly additive value. In integral consideration of the whole nonequilibrium system, its entropy can increase. This problem requires further investigation. One circumstance, however, deserves mentioning. In the literature (e.g. /7/) one can find a statement that in the local-equilibrium distribution the entropy can not increase. In our opinion, this statement is not true and is based on the fact that energy of interaction with the environment of the considered part of the system has not been properly taken into account. Moreover, from the above consideration it should be clear that for the mechanical substantiation of the second law of thermodynamics, there is no need in a non- physical assumption on discrepancy of calculated and measured trajectories, which was made in a widely cited paper by N. S. Krylov /8/.

Devoid of grounds is also the conclusion made by J. Prigogine in paper /9/ concerning the fact that in thermodynamic systems the basic concepts of classical and quantum mechanics cease to conform to experimental data.

## On creation of an exact microscopic heat theory.

By condition of distribution function normalization
We get
$$e^{-\frac{F}{\Theta}} = \int e^{-\frac{E}{\Theta}} d\Gamma$$

Having differentiated both parts with respect to t, taking into account that E depends on environmental particles' coordinates,
$$e^{-\frac{F}{\Theta}} \frac{dF}{dt} \frac{1}{\Theta} = \int e^{-\frac{E}{\Theta}} \frac{dE}{dt} \frac{1}{\Theta} d\Gamma$$

we multiply by $e^{\frac{F}{\Theta}}$
$$\frac{dF}{dt} = \int \frac{dE}{dt} e^{\frac{F-E}{\Theta}} d\Gamma = \left\langle \frac{dE}{dt} \right\rangle = -\langle F_k \rangle V_k$$

As entropy
$$S = -\langle ln\rho_n \rangle = -\int \frac{F-E}{\Theta} e^{\frac{F-E}{\Theta}} d\Gamma = \frac{1}{\Theta}(\langle E \rangle - F)$$

Then
$$\frac{dS}{dt} = \frac{1}{\Theta}\left( \frac{d\langle E \rangle}{dt} - \left\langle \frac{dE}{dt} \right\rangle \right)$$

The same formula but it was obtained according to information theory, and it was cited in a well-known book /10/.

Let us find a time derivative from mean energy
$$\frac{d\langle E \rangle}{dt} = \frac{d}{dt}\int E e^{\frac{F-E}{\Theta}} d\Gamma = \int \frac{dE}{dt} e^{\frac{F-E}{\Theta}} + \frac{1}{\Theta}\int E \frac{dF}{dt} e^{\frac{F-E}{\Theta}} - \frac{1}{\Theta}\int E \frac{dE}{dt} e^{\frac{F-E}{\Theta}} =$$
$$= \left\langle \frac{dE}{dt} \right\rangle + \frac{1}{\Theta}\langle E \rangle\left\langle \frac{dF}{dt} \right\rangle - \frac{1}{\Theta}\left\langle E \frac{dE}{dt} \right\rangle$$

Hence
$$\frac{dS}{dt} = \frac{1}{\Theta^2}\left(\langle E\rangle\left\langle\frac{dE}{dt}\right\rangle - \left\langle E\frac{dE}{dt}\right\rangle\right) = \frac{1}{\Theta^2}\left(\left\langle E\sum_k F_k\right\rangle V_k - \langle E\rangle\sum_k \langle F_k\rangle V_k\right)$$

The obtained formula contains both correlative function and environmental particle velocity. We cannot perform environmental ensemble averaging, unless the system is fully equilibrium as previously mentioned about nonergodicity of nonstationary process. Apparently in this case we should perform time averaging or use some kind of approximation. As a result of interaction of subsystems with different temperatures or hydrodynamic momentum average time value must be different from zero. This matter needs further research.

## Pair distribution function.

By using subsystems ensemble you can easily deduce a two-particle distribution function for pair interaction systems disregarding edge conditions. Hamiltonian operator of such systems looks like this:

$$H_N = T_N(\boldsymbol{p}_1,\ldots \boldsymbol{p}_N) + \sum_{i=1}^N \sum_{j>i}^N W_{ij}$$

Where $T_N(\boldsymbol{p}_1,\ldots \boldsymbol{p}_N) = \sum_{i=1}^N \frac{\boldsymbol{p}_i^2}{2m_i}$ -kinetic energy

$\sum_{i=1}^N \sum_{j>i}^N W_{ij} = \sum_{i=1}^N \sum_{j>i}^N W(\boldsymbol{R}_i - \boldsymbol{R}_j)$ -potential energy of particles interaction

$\boldsymbol{R}_i$ - coordinates

$\boldsymbol{p}_i$ - momentum of $i$-particle

For convenience's sake we use vector notations in formulas (bold font). For simplicity we also ignored particles' energy in external field.

Let us first construct an ensemble for equilibrium system out of all $N$ particles and find for it a distribution function disregarding edge conditions. This will constitute a regular Gibbs distribution

$$\rho_N = e^{\frac{F-H_N}{\Theta}} \qquad (19)$$

Further it is convenient to make a full Hamiltonian look like this

$$H_N = H_{N-2} + H_w + H_2$$

Where $H_{N-2}$ - a Hamiltonian for $N$-2 particles

$H_w$ - an interaction Hamiltonian for $N$-2 particles with two remaining particles

$$H_2 = \frac{\boldsymbol{p}_{N-1}^2}{2m_{N-1}} + \frac{\boldsymbol{p}_N^2}{2m_N} + W_{N-1,N}$$

Then we construct ensemble for $N$-2 particles and we get by labeling

$$E_{N-2} = H_{N-2} + H_W$$

conditional distribution function for ($N$-2) particles

$$\rho_{N-2} = e^{\frac{F(\boldsymbol{R}_{N-1},\boldsymbol{R}_N) - E_{N-2}}{\Theta}} \qquad (20)$$

where $F(\boldsymbol{R}_{N-1},\boldsymbol{R}_N)$ - a normalizing factor depending on coordinates $\boldsymbol{R}_{N-1}$ and $\boldsymbol{R}_N$ particles.

According to the definition of conditional probability
$$\rho_N = \rho_{N-2} f_2(N-1,N) \qquad (21)$$

where $f_2(N-1,N)$ - a required 2-particle distribution function

From formulas ( 21 ) by using ( 19, 20 ) we get

$$f_2 = e^{\frac{F - F(\boldsymbol{R}_{N-1},\boldsymbol{R}_N) - H_2}{\Theta}}$$

We take variation from $F(\boldsymbol{R}_{N-1},\boldsymbol{R}_N)$ in changing the coordinates $\boldsymbol{R}_{N-1}$ and $\boldsymbol{R}_N$.

From formula ( 17 ), assuming $\delta\Theta = 0$ we get

$$\delta F(\boldsymbol{R}_{N-1}, \boldsymbol{R}_N) = -\langle \boldsymbol{F}_{N-1} \rangle \delta \boldsymbol{R}_{N-1} - \langle \boldsymbol{F}_N \rangle \delta \boldsymbol{R}_N$$

Where $\langle \boldsymbol{F}_{N-1} \rangle$ and $\langle \boldsymbol{F}_N \rangle$ - average forces influencing particles *N-1* and *N* on the part of *N-2* other particles. Interaction forces between particles *N-1* and *N* are not taken into account. Since average force influencing the particles far from vessel wall amounts to zero at equilibrium, work and change $F(\boldsymbol{R}_{N-1}, \boldsymbol{R}_N)$ of two particle transition amounts to zero. It does not depend on mutual distance between these two particles either, the energy of their interaction does not count (it is only part in $H_2$).

Therefore $f_2 = e^{\frac{\Psi_2 - H_2}{\Theta}}$ where $\Psi_2 \equiv F - F(\boldsymbol{R}_{N-1}, \boldsymbol{R}_N)$ – it does not depend on $\boldsymbol{R}_{N-1}$ and $\boldsymbol{R}_N$. We may put down configuration part of distribution function taking into account particles symmetry in respect to coordinate permuting for equilibrium system

$$n_2 = \frac{1}{Q_2} e^{-\frac{W_{1,2}}{\Theta}}$$

Where $Q_2 = \int e^{-\frac{W_{1,2}}{\Theta}} d\boldsymbol{R}_1 d\boldsymbol{R}_2$

i.e. two-particle distribution function known as zero-order approximation of BBGKY hierarchy is the exact pair distribution function for equilibrium system disregarding edge conditions.

Many manuals on statistical mechanics contain a derivation of hydrodynamical equations from Boltzmann's equation. Let us consider it in detail, following the work /3/. Boltzmann's equation has the form

$$\frac{\partial f}{\partial t} + \sum_{\kappa=1}^{3} \left( \frac{\partial f}{\partial x_\alpha} \dot{x}_\alpha + \frac{\partial f}{\partial v_\alpha} F_\alpha \right) = J_c \quad (22)$$

where $f = f(x_\alpha, v_\alpha, t)$ is one-particle distribution function, $x_\alpha$ are the particle coordinates, $v_\alpha$ are velocities, $F_\alpha$ are components of a volume force affecting the particle. $J_c$ is the collision integral, the particular form of which is not very important for us. Now essential is just the fact that in this term, and only in it, all the interactions of the particle under consideration with the environment are taken into account.
The left-hand side of (22) represents a one-particle Liouville's equation for a molecule of the ideal gas. Equation (22) is transformed in a standard manner /3/ to Maxwell's transfer equation containing an arbitrary function $f$ of coordinates, time and velocities.

In the cases $f = m$, $f = mv_\alpha$, $f = \frac{mv_\alpha^2}{2}$ the right-hand side of the transfer equation, which takes the particle interaction into account, vanishes /3/. In this connection we shall obtain the same equation that we would obtain if we initially began the derivation from the one-particle Liouville's equation in which account has been taken of only volume forces and by no means particle interaction forces have been taken into account, which being "dissipative" in averaging. Therefore, hydrodynamic equations obtained from Boltzmann's equation are those for ideal liquid. Accordingly, an expression for viscous stress tensor used in this derivation is formal and inexact. Also incorrect, in our opinion, is the conclusion based on this definition, that in an equilibrium state all the gases are non-viscous. One could equally well allege that masses of all the bodies at rest are equal to zero.

## Stochasticity

In the literature on statistical mechanics the statements are often encountered that, in principle, for a thermodynamic system one could determine coordinates and momenta of all the particles and calculate their trajectories with any accuracy, had one an appropriate computer. But in this case a very simple fact is missed that there are no isolated systems in reality. To predict trajectories of a particle group for a long period, one must know coordinates and momenta of particles of the whole universe. It was already evident for Laplace, but it is still ignored in modern works. The reason for it, in my opinion, is well-known Liouville's theorem about conservation of phase volume in Hamiltonian systems. In fact, any Hamiltonian system is a subsystem and it interacts with the environment. The whole universe is an exception, which interacts with nothing and conserves its phase volume (naturally, when considering in the context of classical mechanics). It is for this reason that a strictly deterministic description is not applicable for Hamiltonian systems, the Poincaré's recurrence theorem being completely unusable.

Indeed, the theorem proving rests on two statements: conservation of a phase volume and closure of the system. But "closed" systems, as it should be clear from the above consideration, do not conserve their phase volume. The universe as a whole, conserves its phase volume but is not closed. Therefore in the real world, the applicability conditions of the Poincaré's recurrence theorem are met nowhere and the recurrence time does not exist indeed.

Erroneous is the opinion that this time for thermodynamic systems is finite though large, the time simply does not exist. Evolution in the real world passes unidirectionally.

The preceding consideration allows one to make a bridge between classical and quantum mechanics. Indeed, should we take one particle, even knowing its coordinates and momenta, we could not predict its trajectory! This is well illustrated by the Brownian motion as an example. Even in the framework of classical mechanics any description is statistical. The source of stochasticity is our lack of knowledge. We always have hidden parameters. And, in our opinion, there is no need to seek a source of statistical description in the systems non-linearity as it is done in current literature /4/

## To problem of irreversibility.

It is well known that laws of mechanics are invariant with regard to time reversal $t \rightarrow -t$, whereas the equations of macroscopical physics do not possess this property. Thus, for example, heat always goes from a hotter body to less heated one, rather than conversely. Many people consider this fact mysterious, demanding the additional hypotheses or statements to explain it.

Actually in irreversibility of macroscopical physics equations there is nothing more mysterious than, for example, in the statement that the set of particles with positive momentum components Ox moves to the right along the X-axis though, in principle, the movement is possible in both directions of the axis. In other words, the irreversibility is caused by the fact of existence of initial conditions itself, no matter if they are "good" or "bad", if they are established precisely or inaccurately. And another example: we can uniquely determine a time direction looking through videotape recording of archery though here we deal with only one "particle" - an arrow. In macroscopic physics the temperature, for example, can play the role of initial conditions. As it is clear from previous discussion, assigning temperature defines the function of momenta distribution and, therefore, the direction of kinetic energy transfer at particle collision, that is the direction of heat transfer. It should also be clear that for the full process reversal in any system, for example in a vessel with gas with a temperature gradient, it would be necessary to reverse not only particles momenta of the system under consideration, but also momenta of all other particles in the universe including those the observer consists of. It would automatically result in sign reversal of some coefficients in such equations as thermal conductivity equation, diffusion equation, and so on. Hence it is clear that no absolute reversibility exists.